\documentstyle[aps,prl,multicol]{revtex}
\input{epsf}
\begin{document}
\draft

\title{Spontaneous Emission in Chaotic Cavities}

\author{T.~Sh.~Misirpashaev$^{a,b}$, P.~W.~Brouwer$^a$, 
and C.~W.~J.~Beenakker$^a$}
\address{$^a$Instituut-Lorentz, Leiden University,
                 P.O. Box 9506, 2300 RA Leiden, The Netherlands\\
$^b$Landau Institute for Theoretical Physics, 2 Kosygin Street,
Moscow 117334, Russia}
\date{June 25, 1997}
\maketitle

\begin{abstract}
The spontaneous emission rate $\Gamma$ of a two-level atom inside 
a chaotic cavity fluctuates strongly from one point to another 
because of fluctuations in the local density of modes. 
For a cavity with perfectly conducting 
walls and an opening containing $N$ wavechannels,  
the distribution of $\Gamma$ is given by 
$P(\Gamma)\propto \Gamma^{N/2-1}(\Gamma+\Gamma_0)^{-N-1}$, where
$\Gamma_0$ is the free-space rate. For small $N$ the most probable
value of $\Gamma$ is much smaller than the mean value $\Gamma_0$.
\end{abstract}

\begin{multicols}{2}
\narrowtext

The modification of the rate of spontaneous emission in a cavity 
has been a subject of extensive research
\cite{Kle81,Goy83,Gab85,Jhe87,Hei87,DeM87,Bar96,OPM}. It was 
shown that the cavity can both enhance and inhibit the spontaneous 
emission at microwave and optical frequencies. The effect is due to a
modification by the environment of the local density of modes 
at the position of the radiating atom. 
The efforts were concentrated on
the fabrication of cavities of prescribed regular shape, the atoms 
being kept close to nodes or antinodes of the field patterns of the 
cavity modes. 

What can be said if the shape of the cavity is not regular and
the exact position of the atom is unknown? Irregular cavities have
a complicated ``chaotic'' field pattern, and it becomes difficult to
state whether the spontaneous emission rate $\Gamma$
of a particular atom  is increased or
decreased with respect to the free-space rate 
$\Gamma_0=d^2\omega_0^3/3\pi\epsilon_0\hbar c^3$
(corresponding to an electric dipole transition with 
 moment $d$, frequency $\omega_0$).
Nevertheless, a precise statement can be made about the
statistical distribution of $\Gamma$. The distribution is universal,
i.e.\ independent of the shape or size of the cavity, provided it
is chaotic. 

\begin{figure}
\hspace*{32truemm}
\epsfxsize=0.25\hsize
\epsfysize=0.05\vsize
\epsffile{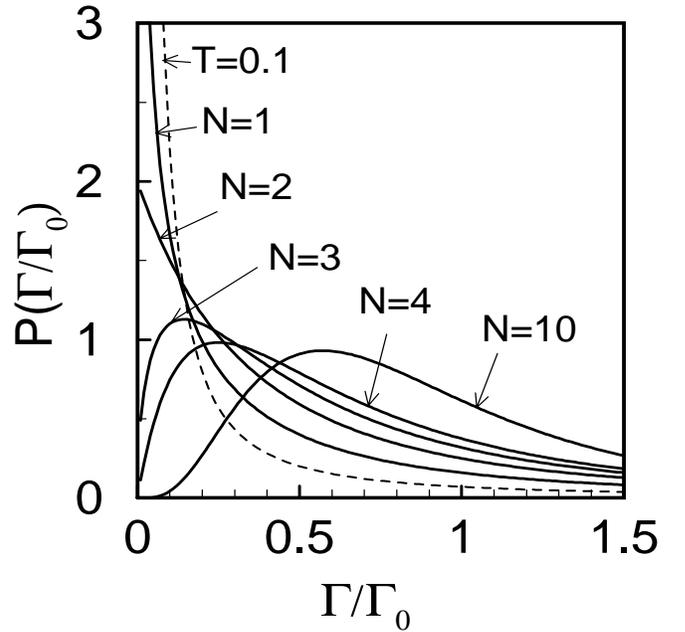}
\vglue 65truemm
\caption{Probability distribution of the spontaneous emission
rate $\Gamma$ (normalized by the free-space rate $\Gamma_0$),
as given by Eq.~(\protect\ref{P_N}) for several values of the number
$N$ of wavechannels transmitted by the hole in the cavity. 
For $N\geq2$ the distribution reaches its maximum at a rate 
$\Gamma=\Gamma_0(N-2)/(N+4)$ that is smaller than the mean value 
$\Gamma_0$. 
The variance of $\Gamma$ diverges for $N\leq2$ and equals
$4\Gamma_0^2/(N-2)$ for larger $N$. The dashed curve is 
the result~(\protect\ref{P_T}) for a hole
much smaller than a wavelength (transmittance $T=0.1$).}
\label{f1}\end{figure}

A chaotic cavity is large compared to the wavelength 
$\lambda_0=2\pi c/\omega_0$,
and has a shape such that the light is scattered uniformly 
in phase space.
(In a circular or cubic cavity, chaotic behavior may still occur
because of diffuse boundary scattering or due to randomly placed
scattering centers.) The only parameter which enters 
the distribution of $\Gamma/\Gamma_0$
is the strength of the coupling of the cavity modes to the 
outside world.
We assume that the coupling is via a hole that is small 
compared to the size
of the cavity and transmits a total of $N$ wavechannels. 
(For a hole of area $A$, $N\approx 2\pi A/\lambda_0^2$.)
Our result for the distribution of $\Gamma$
takes the universal form 
\begin{equation}
P(\Gamma) \propto {\Gamma^{N/2-1}\over(\Gamma+\Gamma_0)^{N+1}},
\label{P_N}\end{equation}
shown in Fig.~\ref{f1} for several values of $N$. 
The distribution eventually becomes narrow and Gaussian for $N \gg 1$, 
while it is still broad and strongly non-Gaussian for 
$N$ as large as 10.
The mean value of $\Gamma$ equals $\Gamma_0$ but the most probable
value is smaller than $\Gamma_0$.
\vglue 5truemm
As a possible experimental setup, one can imagine an array of cavities,
each containing a few excited atoms, or a single cavity containing many 
excited atoms (widely separated so that they decay independently). 
The array of cavities might occur naturally in a porous material.  
Let $n(t)$ be the number of atoms that has not
decayed by the time $t$. The fraction $n(t)/n(0)$ is the Laplace 
transform $\int_0^\infty\!d\,\Gamma P(\Gamma) \exp(-\Gamma t)$ 
of the distribution~(\ref{P_N}), which is a confluent hypergeometric
function. 
A time-resolved measurement of the emitted
intensity yields $n(t)$ and thereby  
the probability distribution $P(\Gamma)$. Fluctuations of the 
spontaneous emission
rate give rise to an algebraic decay $n(t)\propto t^{-N/2}$  
for large $t$, instead 
of the usual exponential decay $\propto\exp(-\Gamma_0 t)$.     

We proceed with the derivation of Eq.~(\ref{P_N}).
We assume that the system is in the perturbative regime \cite{Har92},
so that the rate of spontaneous emission 
is given by the Fermi golden rule,
\begin{equation}
\Gamma=\frac{2}{\hbar^2}{\rm Im}\,\sum_\mu
\frac{\langle f_\mu^L|{\vec d}\cdot{\vec E}|0\rangle
\langle 0|{\vec d}\cdot{\vec E}| f_\mu^R\rangle}
{\omega_\mu-\omega_0-i\gamma_\mu/2}.
\label{Golden}
\end{equation}
Here $|0\rangle$ is the initial state (excited atom $+$ no photons) and
$| f_\mu^{L,R}\rangle$ is the 
final state (atom in the ground state $+$ one photon in mode $\mu$ with
frequency $\omega_\mu$, broadening $\gamma_\mu$). 
The index $L$ or $R$ refers to left and right 
eigenfunctions of the Maxwell equations in 
the open cavity, which form a biorthogonal set of modes.  
The conditions for the validity of perturbation theory
will be discussed later.

Eq.~(\ref{Golden}) can be rewritten
in terms of the local density of modes at the position 
$\vec r$ of the atom,
\begin{eqnarray}
&&\Gamma=\frac{\pi\omega_0d^2}{\hbar\epsilon_0}\rho(\vec r,\omega_0), \\  
&&\rho(\vec r,\omega)=\frac{1}{\pi}{\rm Im}\, \sum_\mu
\frac{E_\mu^{L*}(\vec r) E_\mu^R(\vec r)}{\omega_\mu-\omega-i\gamma_\mu/2},
\end{eqnarray}
where $E_\mu^{L,R}$ is the component along $\vec d$ of the 
electric field in left or right mode $\mu$.
We consider an almost empty cavity without any dispersive or absorptive
medium inside, in which case the distinction between the total
and radiative density of modes \cite{Lag96} is irrelevant.

For a statistical description we study an ensemble of chaotic cavities
with the same volume ${\cal V}$ and
small variations in shape. The average density of modes
$\langle \rho(\vec r,\omega_0)\rangle\equiv\rho_0
=\omega_0^2/3\pi^2c^3$  corresponds to the average rate $\Gamma_0$.
Our aim is to find the probability distribution of $\rho$. 
In Refs.~\onlinecite{Efe93,Bee94,Tan96} this distribution 
was obtained under the assumption that the broadening $\gamma_\mu$
was the same for all modes and all cavities. In our problem,
the broadening is different for each
mode and each cavity, and the distribution turns out to be 
entirely different.  

According to the universality hypothesis of chaotic systems,
the statistical distribution of $\rho$ can be described by the 
random-matrix theory of chaotic scattering \cite{VWZ}. 
 Starting point is the expression of
the $N \times N$ scattering matrix $S$ in terms
of an $M\times M$ real symmetric matrix $H$ (representing 
the discretized Helmholtz operator of the closed cavity) and an 
$M\times N$ coupling matrix $W$, 
 
\begin{equation}   S(\omega) = 1
 - 2\pi iW^\dagger(\omega - H + i\pi WW^\dagger)^{-1} W.
\label{SK}
\end{equation}
The matrix $H$ is taken from the Gaussian orthogonal 
ensemble of random-matrix theory,
\begin{equation}
  P(H) \propto 
\exp\left[-(\pi\rho_0{\cal V})^2 \mbox{tr}\, H^2/4 M\right].
\end{equation}
The limit $M\to\infty$ is taken at the end of the calculation.
The coupling matrix $W$ has elements 
$W_{mn} = (M/\rho_0{\cal V})^{1/2}\pi^{-1}\delta_{mn}$. 

The local density of modes is obtained from a diagonal element of 
the Green function $G(\omega)=(\omega-H+i\pi WW^\dagger)^{-1}$,
\begin{equation}
\rho(\vec r_m,\omega)=-(M/\pi {\cal V}){\rm Im}\,G_{mm}(\omega),
\label{rhoasG}\end{equation}
where $\vec r_m$ is the point in space associated with the index $m$.
Because of the orthogonal invariance of 
$P(H)$, the distribution of $\rho$ is independent of $m$.
Using Eq.~(\ref{SK}), we can rewrite Eq.~(\ref{rhoasG}) in terms of
the scattering matrix,
\begin{equation}
\rho={M\over 2 \pi {\cal V}} i\, \mbox{tr}\, 
    S^{\dagger} \partial S / \partial H_{mm}.
\label{rhoBut}\end{equation}
This representation of the local density of modes is the matrix
analogue of the relationship \cite{But93} between the local density 
of electronic
states and the functional derivative of the scattering matrix 
with respect to the local electrostatic potential,
$\rho(\vec r) = (i/2\pi) \mbox{tr}\, S^\dagger 
  \delta S/\delta V(\vec r)$.

The matrix  $S^{\dagger} \partial S / \partial H_{mm}$
is  closely related to the matrix

\begin{equation}
Q=-iS^\dagger \partial S/\partial\omega,
\end{equation}
known as the Wigner-Smith time-delay matrix \cite{WignerSmith}. 
Namely, in view of Eq.~(\ref{SK}) we have 
\begin{equation}
i\, \mbox{tr}\,  S^{\dagger} \partial S / \partial H_{mm} = 
(AA^\dagger)_{mm},
\qquad Q=A^\dagger A,
\end{equation}
where $A=(2\pi)^{1/2} GW$.
Since $A$ is an $M \times N$ matrix, the product $A A^{\dagger}$ has 
$M - N$ zero eigenvalues. The remaining $N$ nonzero eigenvalues 
are the same as the 
eigenvalues of $Q$, which are the so-called proper delay times 
\cite{Fed97}
 $\tau_1, \ldots, \tau_N$. Their statistical distribution is 
known \cite{Bro97-2},
\begin{equation}
  P(\tau_1,\ldots,\tau_N) \propto
  \prod_{i<j} |\tau_{i} - \tau_{j}|
  \prod_k \tau_{k}^{-3 N/2 - 1}
  e^{-{\pi\rho_0{\cal V} / \tau_k}}. \label{eq:TauDistr}
\end{equation}
For the local density of modes~(\ref{rhoBut}), this implies that
\begin{equation}
  \rho = {M \over 2 \pi {\cal V}}
  \sum_{j=1}^{N} u_{j}^2 \tau_j, \label{eq:Usum}
\end{equation}
where $u_{j}$ is the $j$-th element of the eigenvector 
of $AA^\dagger$
corresponding to the eigenvalue $\tau_j$. In the limit $M \to \infty$, 
the distribution of the vector $\vec u$ is Gaussian, $P(\vec u) 
\propto \exp(-\case{1}{2}M|\vec u|^2)$.

Eq.~(\ref{eq:Usum}), together with the distribution~(\ref{eq:TauDistr}) 
of the 
$\tau_j$'s and the Gaussian distribution of the $u_j$'s, completely 
determines the distribution of $\rho$ and hence of $\Gamma$.  
We replace the integration over the $\tau_j$'s by the 
integration over all elements of an 
arbitrary real $N \times N$ matrix $B$ such that the $\tau_j$'s 
are eigenvalues of 
$(B B^{\dagger})^{-1}$. The matrix $B$ has distribution \cite{Bro97-2}
$P(B)\propto\exp(-\pi\rho_0{\cal V}\,\mbox{tr}\,BB^\dagger) 
(\mbox{det}\,BB^\dagger)^{(N+1)/2}$.  
Using the dimensionless variable 
$x = \rho/\rho_0 = \Gamma/\Gamma_0$ and
properly rescaling $\vec u$, $B$, 
 the integral for the distribution becomes
\begin{eqnarray}
  P(x) &\propto& \int\!\!d\vec u \int\!\! dB\,
    e^{-\mbox{tr}\,BB^{\dagger}-|\vec u|^2}
    \nonumber \\ && \mbox{} \times
    \det(BB^\dagger)^{N+1\over 2}\delta\left(x-|B^{-1}\vec u|^2\right).
\label{Buforx}\end{eqnarray}
We first compute the distribution of the vector 
$\vec v = B^{-1} \vec u$,
which is given by Eq.~(\ref{Buforx}) with the delta function replaced
by $\delta(\vec v-B^{-1}\vec u)$. 
The result is $P(\vec v) \propto (1 + |\vec v|^2)^{-N-1}$.
Due to rotational invariance of the Gaussian distribution for $\vec u$,
the distributions of $x$ and $|\vec v|^2$ are the same. Hence
$P(x)=\int d\vec v\,P(\vec v) \delta(x-|\vec v|^2)
\propto x^{N/2-1}(1+x)^{-N-1}$.
This is the result~(\ref{P_N}) announced in the introduction 
and plotted in Fig.~\ref{f1}. It decreases monotonically for $N\leq2$, 
and has a maximum at non-zero $\Gamma$ for larger $N$.

This calculation holds for the so-called orthogonal symmetry
class (symmetry index $\beta=1$), relevant for optical systems
with time-reversal symmetry.
 The local density of states for systems with broken 
time-reversal symmetry 
(unitary class, $\beta=2$) or with broken spin-rotational symmetry 
(symplectic class, $\beta=4$) is relevant in condensed matter physics.
We have repeated our calculations for 
$\beta=2,4$ and found $P_N^{(\beta)}=P_{\beta N}^{(1)}(x)$, 
with $P^{(1)}(x)$ given by Eq.~(\ref{P_N}). 

So far we have assumed that the hole in the cavity fully transmits 
at least one wavechannel, so that the transmittance $T$ of the hole 
(the ratio of the transmitted and incident power)
is $\geq1$. If the hole
is smaller than a wavelength, then $T$ becomes $<1$. 
The scattering matrix $S(T)$ of the cavity 
coupled by a hole
with transmittance $T<1$ can be expressed in terms of 
the scattering matrix $S|_{T=1}$,
\begin{equation}
  S(T)=\frac{S|_{T=1}+\sqrt{1-T}}{1+S|_{T=1}\sqrt{1-T}}.
\end{equation} 
To find the distribution of the local density of modes, 
we start from Eq.~(\ref{rhoBut}) with $S$ replaced 
by $S(T)$, repeat similar steps and average over 
$S|_{T=1}=e^{i\phi}$ at the end.
The result is
\begin{eqnarray}
  P(x) &=& \frac{2}{\pi^2\sqrt{xT}} 
    \int_0^\pi\, d\phi \nonumber \\ && \mbox{} \times 
    \frac{\sqrt{2-T+2\sqrt{1-T}\cos\phi}}
{[1+x(2-T+2\sqrt{1-T}\cos\phi)/T]^2},
  \label{P_T}
\end{eqnarray}
plotted also in Fig.~\ref{f1} (dashed line, for $T=0.1$). 
It decreases monotonically 
for any $T<1$.

The variance $\langle (\Gamma-\Gamma_0)^2\rangle$ diverges if
$N\leq2$ but the divergency 
is removed when we take into account
the condition of applicability of the Fermi golden 
rule~(\ref{Golden}).
The perturbative treatment is valid as long as the decay rate $\Gamma$
of the excited atom
remains smaller than the width $\gamma_\mu$ of the cavity
modes contributing
to the decay. Estimating the width
of the main contributing mode as $1/\rho{\cal V}=
\Gamma_0/\Gamma\rho_0{\cal V}$, 
we get a condition 
$\Gamma\ll (\Gamma_0/\rho_0{\cal V})^{1/2}$. 
Therefore, any divergent contribution of the large-$\Gamma$ tail 
should be cutoff at $\Gamma\simeq(\Gamma_0/\rho_0{\cal V})^{1/2}$.
The weight of the tail is negligibly 
small provided 
$(\Gamma_0/\rho_0{\cal V})^{1/2}\gg\Gamma_0$, hence if 
$\Gamma_0\rho_0{\cal V}=d^2\omega_0^5 
{\cal V}/9\pi^3\epsilon_0\hbar c^6\ll1$.
To estimate this parameter, we
write $d=zea_B$ ($a_B$ is the Bohr radius), 
$\omega_0=2\pi c/\lambda_0$,
${\cal V} = L^3$. Then $\Gamma_0\rho_0{\cal V}
\approx 3.21 z^2 a_B^2 L^3/\lambda_0^5$
is close to 1 for $z=0.17$, $L=0.53$~mm, $\lambda_0=530$~nm. We can
get large room for applicability of Eqs.~(\ref{P_N}), (\ref{P_T}) by 
going to weaker (possibly magnetic) dipoles, 
smaller cavities, or larger
(possibly microwave) wavelengths.  

We conclude with a comparison 
with previous work on the local density of states in chaotic cavities
\cite{Efe93,Bee94,Tan96}. That work was motivated by 
different physical applications
(Knight shift in NMR or optical absorption). 
Our application is in a sense dual to that of 
Ref.~\onlinecite{Tan96}, where 
complicated electronic states interact with simple radiation states. 
Instead, we have the simplest possible electronic 
system---a two level atom---and a complicated 
structure of radiation modes.
In Refs.~\onlinecite{Efe93,Bee94,Tan96} 
it was assumed that the cavity 
was coupled to the outside via a tunnel barrier of large area. 
In this case
statistical fluctuations in the broadening of the levels $\gamma_\mu$
(from level to level and from cavity to cavity) can be ignored. In the 
case of a relatively small opening, considered here, fluctuations 
of the $\gamma_\mu$'s are essential. 
The resulting distribution~(\ref{P_N}) of the local 
density of modes turns out to be very simple,
compared with the result of Ref.~\onlinecite{Tan96} 
(involving a five-fold
integral in the case of unbroken time-reversal symmetry).
We obtained our result within the framework of
random-matrix theory.
It would be interesting to see if it 
can be reproduced using the supersymmetry technique of 
Refs.~\onlinecite{Efe93,Tan96}.

This work was supported by the Nederlandse Organisatie voor
Wetenschappelijk
Onderzoek (NWO) and the Stichting voor Fundamenteel Onderzoek der
 Materie (FOM).

\end{multicols}

\end{document}